\documentstyle[12pt,cite,epsf]{article}                                        
\catcode`\@=11                                                       
\begin{document}                                                     

\renewcommand{\theequation}{\thesection.\arabic{equation}}        
\newcommand{\mysection}[1]{\setcounter{equation}{0}\section{#1}}  

\def\1{{\bf 1}}
\def\Z{{\bf Z}}
\def\ee{\end{equation}}
\def\be{\begin{equation}}
\def\l{\label}
\def\dxi{\partial_\xi}
\def\D{{\cal D}}
\def\sin{{\rm sin}}
\def\cos{{\rm cos}}
\def\f{{\bf \Phi}}
\def\v{\varphi}
\def\O{\bf {\cal O}}
\def\C{\bf C}
\def\CP{\bf CP}
\def\e{\rm e}
\def\0{\nonumber}
\def\eea{\end{eqnarray}}
\def\bea{\begin{eqnarray}}
\def\Tr{\rm Tr}
\def\IR{\bf R}
\def\ZZ{\bf Z}
\def\T{\tau}

\begin{titlepage}

\hfill{ULB-TH/02-27}

\vspace{2 cm}

\centerline{{\huge{Matrix String Models for Exact (2,2)}}}
\centerline{{\huge{String Theories in R-R Backgrounds}}}

\vspace{2 cm}

\centerline{Giulio Bonelli\footnote{e-mail address: gbonelli@ulb.ac.be}}
\centerline{Physique Theorique et Mathematique}
\centerline{Universite' Libre de Bruxelles}
\centerline{Campus Plaine C.P. 231; B 1050 Bruxelles, Belgium}
\centerline{and}
\centerline{Theoretische Natuurkunde}
\centerline{Vrije Universiteit Brussel}
\centerline{Pleinlaan, 2; B 1050 Brussel, Belgium}

\vspace{4 cm}

{\it Abstract}:
We formulate matrix string models on a class of exact string backgrounds 
with non constant RR-flux
parameterized by a holomorphic prepotential function
and with manifest (2,2) supersymmetry.
This lifts these string theories to M-theory exposing the non perturbative
string interaction which is studied by generalizing the 
instanton asymptotic expansion, well established in the flat background case, 
to this more general case.
We obtain also a companion matrix model with four manifest 
supersymmetries in eleven dimensions.

\end{titlepage}

\section{Introduction}

The quantization of string theories in curved backgrounds
and the study of their finiteness properties is still an open important problem.
It appeared recently an interesting class of (2,2) pp-wave solutions of type IIB 
\cite{mama} generalizing the original one studied in \cite{metsaev}.
These string theories have been shown to admit a superconformal 
formulation in \cite{bema} and have been shown to be exact (finite)
in \cite{bema,ruts}.

The corresponding (2,2) $\sigma$-model has been studied in the $SU(4)\times U(1)$
formalism \cite{GS}.
Within this framework, the type IIB GS action on a flat background is written in terms of four 
complex chiral (2,2) superfields $X^{+l}$, where $l=1,\dots,4$, as
$$
S_0=\int d^2z d^4\theta X^{+l}X^{-l}
$$
where the only part of the original $SO(8)$ symmetry which is left manifest is in fact 
a $SU(4)\times U(1)$ one.
The action relative to the pp-wave backgrounds is written as the $\sigma$-model action
$$
S^{IIB}_{pp}=\int d^2z \left\{\int d^4\theta X^{+l}X^{-l}
+\int d^2\theta W(X^{+l}) + c.c.\right\}
$$
where $W$ is the holomorphic prepotential. The pp-wave metric and the RR-field
curvature are parameterized by this holomorphic function.
Specifically the ten dimensional metric reads
$$
ds^2_{(10)}=-2dx^+dx^--|\partial W|^2(dx^+)^2+2dx^l d\bar x^l
$$
and the RR-fields $F^{(5)}\sim \partial^2 W$.
In \cite{bema} these models have been shown to admit an exact superconformal formulation
(by a suitable reconstruction of the light-cone degrees of freedom) 
and to be therefore exact string solutions.
Notice that this happens for no-renormalization arguments of the
central charge for $(2,2)$ linear $\sigma$-models under an arbitrary shift 
in the holomorphic prepotential
\footnote{
In principle one might consider more general $\sigma$-models by generalizing the 
kinetic term $ X^{+l}X^{-l}\quad\to\quad
K\left(X^{+l},X^{-l}\right) $, where $K$ is the Kahler potential.
In this letter we will not consider such extensions.}.
An interesting aspect of the exact superconformal formulation is that upon
a change of variables the R-NS formulation can be obtained where the string
interaction is well defined, being the background non dilatonic, 
as the usual genus expansion.
Notice that, if $W(x)=\mu\sum_l x_l^2$, then we obtain the maximal
supersymmetric pp-wave with manifest $SO(4)$ symmetry and constant
RR-flux $F^{(5)}\sim\mu$.

The action is given by
\be
S^{IIB}_{pp}=\frac{1}{\pi}\int d^2z \left\{
\partial_+x^l\partial_-{\bar x}^l + 
\partial_+{\bar x}^l\partial_-x^l 
-|\partial_l W|^2 + {\rm fermions}
\right\}
\ee
We see immediately that in the directions in which the pp-wave effectively
extends ($\partial_l W\not=0$), the string coordinate dynamics is that of 
interacting scalar fields. These have in general a very much different 
spectrum with respect to the coordinates propagating in the directions where 
the pp-wave is not effectively extended ($\partial_l W=0$) which are free
massless bosons. Already in case of constant RR-flux pp-wave ($W$ quadratic)
we find a free massive field theory governing the relevant string coordinates
dynamics. This shows that in general the string spectrum
is different in the pp-wave case versus the flat case and that in 
particular the zero mode spectrum will dramatically change.
This can also be shown by studying the supersymmetric classical 
states of the theory. For the flat case, as it is well known, these are
left/right handed propagating waves which are free to locate the string anywhere
because of the existence of the translational zero-mode.
In general, when a non zero prepotential is switched on, this zero modes tends to
disappear\footnote{It is a very well known fact that massive and massless 
quantum field theories are inequivalent -- i.e. the relative realizations of the 
CCRs are mutually irreducible -- exactly for this reason.}
and the string coordinate classically oscillates around the minima of
$|\partial W|^2$.
For several choices of the holomorphic prepotential $W$, there exists
also solitonic solutions which carry topological charges enriching the 
string spectrum.
Also in these cases, the string coordinate elongates around different
minima of the potential $|\partial W|^2$ (periodic solitons).
For this kind of backgrounds there are still some open issues because of the 
absence of asymptotic particle states, being the metric non asymptotically
flat, and because of the irreducibility of solitonic profile strings to punctures.

There is actually a set of similar theories also for type IIA strings.
These can be obtained as subcases of the ones we reviewed above if
a T-duality can be performed along a spacelike isometric direction.
Because of (2,2) manifest supersymmetry, this 
has to be a complex direction. Therefore, to have it manifest, we have to split 
the four chiral superfields $X^{+ l}$ as $\T$
and $\phi^i$,
where $i=1,2,3$, and consider only prepotentials independent on $\T$, i.e.
$\partial_\T W=0$ that is $W=W(\Phi^i)$. 
This is essentially the same condition of existence of at least one transverse 
spectator flat direction considered in \cite{ruts}.
Notice that this choice rules out 
\footnote{
In the supeconformal formulation \cite{bema} of this class of
theories the light-cone coordinate $X^+$ is not a fundamental field,
but it is obtained by a partial on shell procedure. 
This enables one to perform straightforwardly
changes of variables involving light-cone directions to try to render manifest
other spacelike isometric directions.}
the maximal supersymmetric case $W=\mu\sum_l x_l^2$.

Now we are allowed to perform a T-duality transformation (along one of the 
flat directions corresponding to $\T$). The effect of this T-duality is to transform
the chiral superfield $\T$ in a twisted chiral 
\footnote{See \cite{VR} for details.
For definitions, properties and notation here and in the subsequent part of this letter,
we refer the reader to \cite{notation}.}
one that we call $\Sigma$.

The action for type IIA is then given by
\be
S^{IIA}_{pp}=\int d^2z \left\{\int d^4\theta \left\{-\frac{1}{4}\Sigma\bar\Sigma+ \phi^i\bar\phi^i
\right\} +\int d^2\theta W(\phi^i) + c.c.\right\}
\l{22GSA}\ee
Notice that, in particular, if $W=0$, this action reproduces the type IIA GS action on a flat background.

In principle we could add by hand to (\ref{22GSA}) also a twisted prepotential term
$$\int d^2z d\theta^+d\bar\theta^- U(\Sigma)|_{\theta^-=\bar\theta^+=0} + {\rm h.c.}$$
for an arbitrary holomorphic function $U$ in the twisted chiral field.
This corresponds to switch on an additional background RR 2-form flux along the directions
$+$ and one within the complex directions spanned by the scalar components of $\Sigma$. 
Allowing such couplings one can easily reproduce, by choosing quadratic prepotentials,
a whole set of manifest (2,2) type IIA strings on pp-wave backgrounds with constant RR-fluxes.
We will not study these additional couplings here and just comment about their
effect in the concluding section.

At this point the possibility of lifting to M-theory these backgrounds
can be posed clearly.
Matrix String Theory \cite{etc} (MST) links perturbative string theory
and M(atrix)-theory \cite{bfss} by explaining how the theory of strings is included 
in the latter. In particular,
this realizes type IIA string theory on flat background
as the superconformal fixed point of the (8,8) Super-Yang-Mills theory with gauge group
U(N) at the strong Yang-Mills coupling limit and in the large N regime. 
The interacting structure of perturbative string theory, namely the genus expansion, 
is recovered \cite{MST} as an asymptotic expansion of the gauge theory partition sum around the 
conformal fixed point in the inverse gauge coupling which is then,
accordingly with S-duality, interpreted as the string coupling.
This asymptotic expansion is concretely built as a WKB expansion around certain supersymmetric
instanton configurations whose spectral data encode the relative string Mandelstam diagrams.
Notwithstanding its not competing power in evaluating perturbative string amplitudes (although
it solves positively the old problem with genus proliferation in certain high energy 
regimes of string amplitudes \cite{GHV} and enables the exact evaluation of 
protected couplings and supersymmetric indices allowing interesting duality checks), 
since it models string interactions non-perturbatively,
MST has a strong conceptual importance.
It is therefore crucial, once some string background is studied, to obtain its Matrix String 
Theory counterpart to properly embed that perturbative string theory in M-theory.
This has been done for pp-wave backgrounds with constant R-R flux in \cite{ppmst}
and in \cite{also}.
Here we will discuss this issue for a wide set of (2,2) backgrounds with non constant R-R flux.

To this end, we will first formulate Matrix String Theory on flat backgrounds 
in terms of (2,2) superfields in order to reach a comfortable homogeneous framework.
Then we will show how MST generalizes to the description of the above pp-wave backgrounds.
We study the quantum stability of the strong coupling limit and sketch an asympthotic
expansion of the partition sum in such a regime.
As a byproduct of our analysis we obtain also a novel class of matrix models on eleven dimensional
pp-wave backgrounds with non constant 4-form flux and four manifest supersymmetries.
Comments and open questions are contained in the last section.

\section{Matrix String Theory in (2,2) superfields}

Type IIA Matrix String Theory \cite{etc,MST} can be obtained by reducing the ${\cal N}=1$ D=10 SYM
theory with gauge group $U(N)$ down to two dimensions \cite{sym}. 
In order to write down the MST action in the (2,2) superfield formalism it is 
more useful to perform an intermediate dimensional reduction to ${\cal N}=4$ D=4 SYM
which can be written in ${\cal N}=1$ superfields. In these terms 
its spectrum is given by 
a vector multiplet $\cal A$ and three chiral hypermultiplets $\Xi^i$ in the adjoint 
representation of the $U(N)$ gauge group.
The only additional datum which specifies the action is the prepotential
$L=g \Tr \Xi^1[\Xi^2,\Xi^3]$, where $g$ is the gauge theory coupling constant.

Under dimensional reduction ${\cal N}=1$ susy in D=4 reduces to (2,2) susy in D=2
and in particular we have the following reduction map
$$
{\rm Vector} [{\cal A}] \qquad \to \qquad {\rm Twisted}\quad{\rm Chiral} [\Sigma]
$$ $$
{\rm Chiral} [\Xi^i] \qquad \to \qquad {\rm Chiral} [\Phi^i]
$$
By the above map we obtain the MST action in (2,2) superfields as 
\be
\int d^2z d^4\theta \left(-\frac{1}{4g^2}\Sigma\bar\Sigma + \Phi^i\bar\Phi^i\right)
+ \int d^2z \left[d^2\theta L(\Phi^i) + c.c.\right]
\l{22mst}
\ee
with the prepotential $L=g \Tr \Phi^1\left[\Phi^2,\Phi^3\right]$.
Obviously, the three chiral superfields $\Phi^i$ are still in the adjoint 
representation of the gauge group $U(N)$ and the trace over the gauge indices 
is understood.

Notice that in the (2,2) superfields formalism only a $U(1)\times SU(3)$
subgroup of the original $SO(8)$ R-symmetry is manifest (exactly as it was 
in the previous section for the action (\ref{22GSA}) in the flat case.).

The strong gauge coupling expansion of MST can be performed in this formalism 
directly only for the chiral superfield degrees of freedom.
The chiral superfields extremal values are dictated by the vanishing of the 
first derivative of the prepotential $\partial_{\Phi^i}L=\frac{1}{2}\epsilon_{ijk}
\left[\Phi^j,\Phi^k\right]=0$. This selects them to lay on a common Cartan subalgebra $t$.
As far as $\Sigma$ is concerned, one has to enter its elementary fields content
to find the analogous commutators (see \cite{notation} to check explicit formulas)
which select the twisted chiral superfield to lay on the same Cartan subalgebra $t$.

The strong coupling limit action is then, after the rescaling of the
gauge field to reabsorb the gauge coupling in the covariant derivatives,
\footnote{The quantum exactness of this limiting procedure in guaranteed 
by supersymmetry at one loop where the non-Cartan sector is effectively
integrated out. For more details see \cite{MST} and next section.} 
$$
S=\int d^2z \int d^4\theta \left\{-\frac{1}{4}\Sigma^t\bar\Sigma^t+ {\Phi^t}^i\bar{\Phi^t}^i\right\}
$$
where the suffix $t$ projects the fields on the Cartan component along $t$.
This is the MST version of (\ref{22GSA}) on a flat background
i.e. the $S_N$ symmetric superposition of $N$ copies of the type IIA 
tree level string theory.

\section{MST on (2,2) pp-wave geometries}

The point we are addressing in this letter in an extension of MST, as given in the previous section,
to the exact (2,2) pp-wave geometries that we review in the introduction.

The natural model to study is given by a gauging of the type IIA action (\ref{22GSA})
as
\be
S=\int d^2z \left\{\int d^4\theta \left\{-\frac{1}{4g^2}\Sigma\bar\Sigma+  
\Phi^i\bar\Phi^i
\right\} +\int d^2\theta \left(L(\Phi^i)+\tilde W(\Phi^i)\right) + c.c.\right\}
\l{pp22mst}\ee
where now the superfields are relative to the $U(N)$ gauge theory (we keep the notation of the 
previous section where the trace over the gauge group indices is understood).
$\Sigma$ and $\Phi^i$ are the same $U(N)$ superfields we considered in the previous section
for the flat case.
As far as the definitions of the matrix function $\tilde W$
the natural requirement is that once evaluated on Cartan fields it reproduces
the prepotential in (\ref{22GSA}) as 
$\Tr\tilde W\left({\Phi^t}^i\right)=\sum_m
W\left({\Phi_m}^i\right)$ in an orthonormal basis of $t$.
This requirement specifies these structure function 
to be given by $W(\Phi)$
up to the prescription of the relevant matrix ordering.

The natural ordering is of course the total symmetrization.
In the related context of D-geometry this problem has been 
clarified \cite{dgeom} by showing that if the background satisfies the string
equations, then the total symmetric ordering is the correct one
in order to reproduce the correct open string masses assignments.
Since the string backgrounds we are considering are exact, 
TS-duality with the type IIB D-string picture justifies this ordering.
Notice moreover that these D-geometry arguments have been already 
applied in a similar context \cite{brwy} for matrix string theory
on Kahler backgrounds.
In principle one could also consider more general models
by allowing a non quadratic Kahler potential in (\ref{pp22mst}).
We consider the above arguments 
\footnote{See the conclusions for further arguments related to the preservation
  of the background symmetries.}
enough to justify the total symmetric ordering.

For completeness, we give the bosonic part of the action (\ref{pp22mst})
that is
$$
S_b=\frac{1}{2}\int d^2z\left\{
D_{\bar z}\phi^iD_z\bar\phi^{\bar i}+
D_{z}\phi^iD_{\bar z}\bar\phi^{\bar i}+
D_{\bar z}\sigma D_z\bar\sigma+
D_{z}\sigma D_{\bar z}\bar\sigma+
\right.$$ $$\left.
\bar F^{\bar i}F^i+\frac{1}{2g^2}|F_{z\bar z}|^2
-\frac{g^2}{2}[\sigma,\bar\sigma]^2-\frac{g^2}{2}[\phi^i,\bar\phi^{\bar i}]^2
+g^2[\bar\sigma,\phi^i][\sigma,\bar\phi^{\bar i}]
+g^2[\sigma,\phi^i][\bar\sigma,\bar\phi^{\bar i}]
\right\}
$$ 
where the $F$-term is fixed to
$\bar F^i=\frac{\partial(\tilde W+L)}{\partial\Phi^i}(\phi)$.

Notice that the model we are considering is the dimensional reduction from
four to two dimensions of a generalized Leigh-Strassler deformation \cite{LS} of 
${\cal N}=4$, D=4 super Yang-Mills.

\subsection{The strong gauge coupling limit}

We assume that the moduli in the prepotential
(namely the pp-wave mass) are finite with respect to the rescaling gauge coupling
constant. This means that, as far as the strong gauge coupling analysis 
of the classical potential, we can ignore the 
subleading effects due to the presence of a non null $W$.
Therefore the strong gauge coupling limit still implies
that in that regime only Cartan valued fields survives, the 
complement being suppressed.

Therefore the strong gauge coupling limit of [\ref{pp22mst}] is the 
expected $S_N$ symmetric superposition of copies of [\ref{22GSA}],
that is
\be
S^{\infty}=\int d^2z \left\{\int d^4\theta \left\{
-\frac{1}{4}
\Sigma^t\bar\Sigma^t+  
{\phi^t}^i\bar{\phi^t}^i
\right\} +\int d^2\theta W({\phi^t}^i) + c.c.\right\}
\l{22GSA'}\ee
where we denote by the index $t$ the Cartan component.
From a full quantum field theory viewpoint,
the above discussion has to be implemented by 
performing the path integration over the non Cartan direction fields 
(as in \cite{MST}).
The strong coupling expansion of the gauged $\sigma$-model action reads
$$
S=S^{\infty} + 
\int d^2z \left\{\int d^4\theta \left\{
-\frac{1}{4}
\Sigma^n\bar\Sigma^n
+ {\Phi^n}^i \bar{\Phi^n}^{\bar i}
\right\} +\int d^2\theta \Phi^n[\Phi^t,\Phi^n] + c.c.\right\}
+ O(\frac{1}{\sqrt g})
$$
where the superscript $n$ indicates the projection on the complement of the Cartan
algebra $t$ in the matrix field space and the proper superfields dressing with
$e^{iV^t}[\,\,]e^{-iV^t}$,
where $V^t$ is the Cartan component of the full vector superfield $V$ from
which $\Sigma$ is built as the superfield curvature.
This leads, at lowest order in the inverse gauge coupling, 
to a supersymmetric (trivial) Gaussian path integral, 
fully justifying (\ref{22GSA'}) in the quantum theory.

The MST interpretation and the world-sheet reconstruction from string bits 
apply then to our model exactly as in the flat case.

Notice that we assumed that the background moduli do not interfere with the 
strong gauge coupling limit. It would be interesting to study in detail 
how the above picture changes if some pp-wave mass scales with the gauge coupling.
In general, it is easy to see that in a background of this type, 
one can freeze string oscillation along some transverse directions 
just by switching on an high enough mass parameters.
From the MST model we gave above, string interaction do not spoil this
picture, but the strong gauge coupling limit is possibly interfered
by the scaling pp-wave moduli. 
It would be very interesting to formulate a picture for these phases
of the theory.

\subsection{String Interactions from matrix string instantons}

An interesting property of the (2,2) supersymmetric models that we are studying
is that the supersymmetric stringy instantons encoding the string interaction
diagrams as spectral curves \cite{MST} are still manifest
\footnote{See also \cite{nuovo}, where a constructive method is systematically applied 
to the relevant Hitchin system.}.
This is because of the existence of the two flat directions 
relative to the scalars in the twisted multiplet.
Due to the fact that these scalars do not enter the F-terms,
it is in fact still possible to polarize 
matrix string diagram in those direction. The relevant
instanton is modeled, as is the flat case, by the Hitchin system.
We will find that if we try to polarize instanton
matrix string diagrams in the directions in which the pp-wave extends (namely
the $\phi^i$ directions), we fail due to a general argument which implies the
staticity of supersymmetric saturated solutions involving these fields as active.
In this section we sketch a picture of the reconstruction of the genus
expansion of the partition sum for type IIA strings in the geometries that we discussed so far.

We consider field configurations saturating some of the manifest supersymmetry
of the model at hand. 
As usual we set the fermion fields to zero and study the stability 
of this condition under supersymmetry transformation. This results in a set of 
first order equations for the bosonic fields.

As far as the three chiral multiplets are concerned, we have
$$
\delta\psi^i_+=
\sqrt{2}\epsilon_+F^i
-2\bar\epsilon_+[\bar\sigma,\phi^i]
+i\bar\epsilon_-\sqrt{2}(D_0+D_1)\bar\phi^i
$$ $$
\delta\psi^i_-=
\sqrt{2}\epsilon_-F^i
+2\bar\epsilon_-[\sigma,\phi^i]
-i\bar\epsilon_+\sqrt{2}(D_0-D_1)\phi^i
$$
where $\Phi^i=\phi^i+\theta\cdot\psi^i+\theta^2F^i$, $\sigma$ is the
complex scalar in the twisted multiplet and the relevant $F$-term
is $\bar F^i=\frac{\partial(\tilde W+L)}{\partial\Phi^i}(\phi)$.

As far as the twisted multiplet is concerned, the supersymmetric variations of 
its fermions $\lambda_\pm$ are
$$
\delta\lambda_+=\left(iD-F_{01}-[\sigma,\bar\sigma]\right)
\epsilon_+ +\sqrt{2}
(D_0+D_1)\bar\sigma\epsilon_-
$$ $$
\delta\lambda_-=\left(iD+F_{01}+[\sigma,\bar\sigma]\right)
\epsilon_- +\sqrt{2}
(D_0-D_1)\sigma\epsilon_+
$$
(and analogous for $\delta\bar\lambda_\pm$)
where the relevant $D$-term is $D=-i\sum_i[\phi^i,\bar\phi^i]$.

\paragraph{1/2 BPS matrix string configurations}

From the equations above we see immediately that we can saturate diagonal
supersymmetry (i.e. preserve lets say arbitrary $\epsilon_\pm$) if
\be
F_{01}+[\sigma,\bar\sigma]=0
\quad
(D_0+D_1)\bar\sigma=0
\quad
(D_0-D_1)\sigma=0
\l{prehitch}\ee
while the $\phi^i=x_i\times{\bf 1}_N$ are constant diagonal matrices whose eigenvalues
satisfy the equation $\partial_i W(x)=0$. 
\footnote{
The complete set of BPS equations include $\sum_i[\phi^i,\bar\phi^i]=0$,
$[\sigma,\phi^i]=0$, $[\bar\sigma,\phi^i]=0$, $D_\pm\phi^i=0$
and $\frac{\partial (\tilde W+L)}{\partial\phi^i}=0$.
Assuming $\sigma$ truly active, i.e. assuming that its spectrum has not to be
algebraically constrained, we find the condition above on $\phi$.
For disconnected matrix string configurations, this has to be true just block by
block.}
Upon the relevant Wick rotation, equations (\ref{prehitch}) become the Hitchin 
system \cite{H}
\be
F_{z\bar z}+[\sigma,\bar\sigma]=0
\quad{\rm and}\quad
D_z\sigma=0
\ee
The spectral data classifying its solutions is given by the moduli
space of plane curves in two (complex) dimensions, i.e. the light-cone and 
$\sigma$, of order $N$. 
The generic curve ${\cal S}$ is defined by the $\sigma$-spectral equation
$$
0={\rm det}(\sigma(z)-\sigma{\bf 1}_N).
$$
The way in which these spectral curves represent
the Mandelstam diagrams is discussed in detail in \cite{MST}.

\paragraph{1/2 BPS solitonic string configurations}

We can otherwise saturate the supersymmetry as $\bar\epsilon_\pm=\epsilon_\mp$.
This embeds in the gauged sigma model the usual static BPS solutions (solitons
for appropriate choice of $W$).
We find that $\sigma$ is forced to be passive (i.e. $\sigma\propto {\bf 1}_N$)
and the gauge connection to be flat $F_{01}=0$ (still not trivial on the cylinder).
We are left then with the equations
\be
D_0\phi^i=0
\quad,\quad
\sum_i[\bar\phi^i,\phi^i]=0
\quad{\rm and}\quad
iD_1\phi^i+F^i=0
\l{gsolitons}\ee
Solutions to these equations are long static (solitonic) strings in the $\phi^i$
directions whose bits are modeled on the $S_N$-twisted (soliton) spectrum of the
associated ungauged $\sigma$-model. 

From the supersymmetric variations $\delta\lambda$ and $\delta\psi$
it is not possible to saturate supersymmetry in other ways (up to phase redefinitions
of the preserved supersymmetries).
Namely, it is not possible to obtain instantonic configurations 
with active $\phi^i$s since
only static solutions saturate the Bogomonly bound formula for $F$-terms.

As in the flat case, in the strong coupling limit, the partition sum can
be calculated as a WKB expansion around the Hitchin system solutions.
The Cartan field content, which is left as the effective field spectrum at
strong coupling, lifts 
\footnote{There is an ambiguity regarding the lift of the fermions because of
the need of choosing their spin structure. By mediating over all the
possible inequivalent lifts we reproduce the spin structure sum.}
to the spectral curve ${\cal S}$ of the relevant instanton configuration 
and the action $S^\infty$ can be rewritten as
\be
S^{\infty}=\int_{\cal S} d^2z \left\{\int d^4\theta 
\left\{-\frac{1}{4}\hat\Sigma \hat{\bar\Sigma}+  
\hat{{\phi}^i}\hat{\bar{\phi}^i}
\right\} +\int d^2\theta W(\hat{{\phi}^i}) + c.c.\right\}
\l{22GSA'S}\ee
where we denoted by a hat the lifted fields.
This is the GS type IIA string action on the matrix string worldsheet
plus a decoupled $U(1)$ Maxwell field where the worldsheet metric is 
the Mandelstam one.
The integration over the $U(1)$ gauge field (see again \cite{MST} for details)
produces -- because of the rescaling of fields in the calculation of the strong
coupling limit -- a factor of
$g^{\chi_{\cal S}}$, where $\chi_{\cal S}$ is the Euler characteristic of
${\cal S}$. This
gives the correct perturbative weight to the string amplitude and identifies
the genus counting parameter (i.e. the string coupling) as $g_s=g^{-1}l_s^{-1}$.
All this happens as a consequence of the $(2,2)$ supersymmetry of the model
which preserves the full structure of the gauge supermultiplet fields.

Let us notice that we assumed that the strong gauge coupling 
limit is not interfered by the pp-wave mass scale $\mu$ --
inserting dimensionfull parameters $W=\mu w(\phi/l_s)$, where $l_s$ is the
string scale. This translates to the condition
$\mu g_s<<l_s^{-1}$ which is the range of validity of our analysis.

\section{A related set of matrix models}

The above model admits a lift to an eleven dimension matrix model
on the lifted background geometry.
This can be obtained just by dimensional reduction of the gauged
$\sigma$-model that we considered so far
to a 1+0 matrix quantum mechanics with four manifest supersymmetries.
This matrix model corresponds to M-theory (strongly coupled type IIA)
in a pp-wave background.
This eleven dimensional background can be calculated as
\be
ds^2_{(11)}=-2 dx^+dx^- -|\partial_i W|^2 {dx^+}^2 +
d\phi^id\bar\phi^{\bar i}+(d\sigma^a)^2 
\l{11}\ee
$$
F_{(4)}=dx^+\wedge \omega^{(3)}
$$
where $a=1,2,3$. Here $\omega^{(3)}$ is the harmonic three form satisfying the
supergravity equations, the only non trivial equation being $R_{++}\propto |F_{(4)}|^2$.
This amounts to
$$
\left(\partial_a^2+\partial_j\partial_{\bar j}\right) |\partial_i W|^2
= \partial_j\partial_{\bar j} |\partial_i W|^2
\propto |\omega^{(3)}|^2
$$
Which is solved by the choice
$\omega_{\bar i \bar j k}=c{\epsilon_{\bar i\bar j}}^l\partial_l\partial_k W$, where $c$ is a
normalization numerical constant, and all other components vanish
but the complex conjugate.
This is in fact the 11 dimensional lift of the R-R background field we
considered so far.
Notice the appearing of an explicit $SO(3)$ isometry which rotates the $\{\sigma^a\}$
directions in (\ref{11}).

The matrix model, obtained by dimensional reduction to one dimension,
therefore generalizes to non constant fluxes the matrix models 
elaborated in \cite{CLP,ppmst,Lee}.
The action can be written in components as $S=S_b+S_f$, where
$$
S_b=\int dt\left\{
(\partial_t\sigma^a)^2 + |\partial_t\phi^i|^2
-\frac{g^2}{4}\left[\sigma^a,\sigma^b\right]^2
-\frac{g^2}{4}\left[\sigma^a,\phi^i\right]\left[\sigma^a,\bar\phi^{\bar i}\right]
\right. $$ $$\left.
-\frac{g^2}{2}\left[\phi^i,\bar\phi^{\bar i}\right]^2
+|\frac{g}{2}\epsilon_{ijk}[\phi^j,\phi^k]+\partial_i\tilde W|^2
\right\}
$$
and 
$$
S_f=\int dt\left\{
\frac{i}{2}\psi^t\partial_t\psi-\frac{g}{2}\bar\psi\Gamma^a[\sigma^a,\psi]
-g\bar\psi\Gamma^i[\bar\phi^{\bar i},\psi] + c.c.
-c \bar\psi\Gamma^{\bar i \bar j k}\epsilon_{\bar i \bar j}^{\quad l}
\partial_l\partial_k\tilde W \psi +c.c.
\right\}
$$
The study of the rich BPS spectrum of this set of matrix models deserves 
a deep analysis.

\section{Conclusions and open questions}

In this paper we have worked out a type IIA analog of the (2,2) exact pp-wave 
backgrounds with non constant RR-flux studied in \cite{mama}.
Then we have generalized to this set of pp-wave backgrounds
the Matrix String Theory picture. This has been
tested by showing the quantum stability of the strong coupling regimes 
in which the long strings are generated.
String interaction has also been recovered thanks to the existence
of a flat complex direction.
In the last section, we also obtained a set of new matrix theories
on eleven dimensional pp-wave backgrounds with non constant 4-form flux.

A first comment about string interactions and symmetries is in order.
As matrix string theory gives a nonperturbative (although involved)
definition of interacting string theory, it is interesting to consider
the relation between symmetries of the back-ground and interactions.
Our type IIA background is explicitly invariant under $SO(2)$ (acting on the 
$\sigma$ complex plane) and the $SU(3)$ transformations under which 
$W$ is invariant (up to additional constants).
Let us notice that, since the additional prepotential $L=g\Tr\Phi^1[\Phi^2,\Phi^3]$
is fully $SU(3)$ invariant and since the (trace of the) total symmetrization preserves 
\footnote{This can be shown explicitly as follows.
Let $\gamma\in\Gamma$ be such that $W(\gamma\cdot\phi)=W(\phi)+k_\gamma$, with
$k_\gamma$ such that the group action is well defined and $\Gamma$ a subgroup
of $SU(3)$.
Since the total symmetrization commutes with linear transformations,
we have
$$
\tilde W(\gamma\Phi)
=
\widetilde{W\cdot\gamma}(\Phi)
=
\tilde W(\Phi)+k_\gamma {\bf 1}_N $$
Notice that another ordering choice, not commuting with linear
transformations, would produce an interacting
model which does not preserve background symmetries.} 
all possible $W$ invariances (up to the same constant) in $SU(3)$, we obtain a
picture in which the interacting theory preserves naturally all the background symmetries.
Let us notice that this agrees with the issue raised in \cite{rudi}
where the symmetry of the background pp-wave metric is token
as a guiding principle for the construction of a well defined 
string perturbation theory.

In section 3 we have not added by hand to the action a twisted chiral
prepotential of the type already discussed in the introduction.
The type IIA background in this case would result from a compactification
on a circle of a generalization of the eleven dimensional background 
given in the last section.
The addition of such a twisted prepotential term would not change at all the strong 
coupling analysis
of section 3.1 and would generate additional dielectric couplings 
generalizing the ones already discussed in \cite{ppmst}.
These additional couplings have anyway a drawback consisting of a shift in the 
D-term which changes the BPS equations and possibly the instanton equations 
studied in section 3.2 (which were given by the flat transverse direction). 
This seems to imply the need of a further 
refinement of our model (or of our analysis) for those cases.
Actually, the exactness of the pp-wave backgrounds is proven as 
far as the absence of $\alpha'$ corrections is concerned, but nothing is known
about possible string higher genus corrections due to a still too weak control
on perturbative string interaction on such kind of backgrounds.
The implementation of corrections of this kind would significantly modify our picture, 
because of the relation between the gauge coupling and the string coupling,
in the structure of the gauge sector in the finite gauge coupling regime.
Moreover, the MST realization of the model studied in \cite{HS} should be recovered.
We do not study this very interesting issues here and we leave them for future 
researches.

Let us notice that, as well explained in \cite{ber}, working in 
the light-cone gauge is extremely hard as far as the concrete string
amplitudes calculations are concerned.
Despite that,
it would be very interesting to develop a string bits model
as an effective theory for the matrix string bits in these backgrounds
in order to use it as a possible calculational tool for a comparison 
between states in the interacting string theory on pp-waves
and possible supersymmetric gauge theory duals along the lines of \cite{line}.
Notice that in the flat case, the construction of the DVV vertex is 
fixed by the $SO(8)$ R-symmetry and conformal dimension, while in the 
generic (2,2) pp-wave background we have much less R-symmetry
and therefore, in principle, more possible candidates.

The matrix string models that we have formulated here
can be generalized to include also real Killing potentials if
holomorphic transverse isometries are gauged and subsequently frozen.
Moreover, our model can be generalized to the case in which the transverse 
three complex dimensional space is a non compact orbifold ${\bf C}^3/\Gamma$,
where $\Gamma$ is a discrete subgroup of $SU(3)$ whose action is a symmetry 
(up to an additional constant) of the prepotential $W$. 
It would be very interesting to generalize the 
${\cal N}=2$ Landau-Ginsburg techniques elaborated in \cite{notation}
to study then blow-ups of these orbifold singularities
and geometric transitions in general.
This leads directly (see also \cite{vafa} for further motivations) to the 
issue concerning if Matrix String Theory 
can effectively improve our understanding of gauge/string dualities
and eventually make it deeper.
A tempting conjectural picture arises by considering the Dijkgraaf-Vafa \cite{DV}
prescription relating the evaluation of exact prepotentials in ${\cal N}=1$
four dimensional gauge theories via matrix models
(see also \cite{DHKS}).
In these terms, gauge/string duality seems related to an extension of the 
IKKT \cite{IKKT} matrix description of the type IIB string theory
in terms of D-instantons. 
This, upon double dimensional oxidation induced by two T-dualities, can be related 
(exactly in the case in which a spectator complex flat direction is present)
to the matrix string picture which is TS--dual to type IIB D-strings.
Therefore, it arises a conjectural picture in which matrix string theory
would be an effective non perturbative link between gauge four dimensional 
theories and string theories.
This kind of path is not unexpected \cite{mayr}. After all, gauge/string correspondence 
is a manifestation of the open/closed string duality
and the effective duality chain advocated in \cite{DV}
relays on their topological versions at planar/tree level respectively.

\vspace{1 cm}

\noindent
{\bf Acknowlodgements}: 
I would like to thank M.Bertolini, L.Bonora, A.Hammou, J.F.Morales and R.Russo for discussions.
I would like also to thank the organizers of the workshop 
``The quantum structure of spacetime and the geometric nature of fundamental
interactions'' in Leuven, during which part of this work has been developed, for the 
stimulating atmosphere they have created.
Work supported by the European Community's Human Potential
Programme under contract HPRN-CT-2000-00131 Quantum Spacetime
in which G.B. is associated to Leuven.

\end{document}